# A Model of Classical and Quantum Measurement

**Charles Francis**

**Abstract:**

   We take the view that physical quantities are values generated by processes in measurement, not pre-existent objective quantities, and that a measurement result is strictly a product of the apparatus and the subject of the measurement. We habitually make an inaccurate statements when we speak of the measurement of a quantity by an apparatus. These statements can be formalised as a many valued logic with the structure of a vector space with a hermitian form in such a way as to generate probabilities in the results of measurement. The difference between this and classical probability theory is that we are not finding probabilities generated by unknown variables, but probabilities generated by unknown structure. We thus interpret quantum logic as the application of complex truth values to statements in an inaccurate language, and find that the properties of vector space hold for approximate measurement as well as for measurement of optimal accuracy, suggesting that Planck's constant governs the scale of the fundamental structures of matter.

Charles Francis
Clef Digital Systems Ltd
Lluest, Neuaddlwyd
Lampeter
Ceredigion
SA48 7RG

# A Model of Classical and Quantum Measurement

## 1  Measurement

The theory of "fuzzy" or "unsharp" measurement has been developed by Progovecki [1], Holevo [2], Busch and Lahti [3], and others, to deal with inaccuracies in the measurement process. In a recent paper Sturzu proposed that unsharp measurement may be related to the interpretation of quantum mechanics [4]. In this paper we provide a rationale for measurement in which this is in fact the case, and which unifies the ideas of classical, quantum and unsharp measurement.

As generally described in text books, in classical measurement it is assumed that a property of nature exists with a numerical value, and that the object of measurement is to determine it [5]. We will call this the notional value of the measured property. The concept of a notional value clearly breaks down in quantum mechanics in which properties only exist when the state is an eigenstate of an observable operator [6]. Since quantum effects limit the accuracy of any classical process, the implication must be that even in classical measurement the notional value does not exist to arbitrary accuracy, and that we should revise our assumptions about measurement.

In this paper we regard measurement as a generic physical process yielding a value as a result of the process. It is not in general the determination of a pre-existent value, but the generation of a value from a physical process. In classical measurement, the 'error' is the difference between the measured value and the notional value. But if we reject that notion and define the quantity to be the measured value, there is no error, merely a distribution of actual values. Then physical laws (both classical and quantum) do not predict the behaviour of notionally true values, but rather the behaviour of distributions of measured results. The notional value of a quantity is often assumed to be a real quantity, measured to an accuracy within a finite range, but there is no way to justify this assumption. The results of measurement are always integers in units of the finite resolution of the apparatus.

The existence of a measurable value does not imply that a measurement has actually taken place. The value is assumed to be property of the physical processes which take place in measurement, and exists whenever a suitable process takes place, whether or not measurement by an observer is involved. Thus a table has length as a result of the physical processes within the structure of the table even when it is not measured. But unlike the notional value of length, the length of the table is not an exact numerical value, but a distribution of values dependent on the actual structure of the table.

When we measure the length of a table using a ruler, we compare two structures of matter (the table and the ruler) both of which have the numerical property of length, and neither of which is more fundamental than the other. We do not measure the length of the table directly. We actually measure the length of the ruler to one of its marks and assign the value found to the length of the table. Although there is a possibility of error in the assignment, it is possible in principle to exercise sufficient care that such error is eliminated up to the accuracy of the marks on the ruler. This paper will not be concerned with this error, but with the distributions of results obtained by quantum and unsharp measurements, on the assumption that human error is eliminated.

As observed by de Broglie, all measurements can be reduced to measurements of position. For example a classical measurement of velocity may be reduced to a time trial over a measured distance, and a typical quantum measurement involves plotting the path of a charged particle in a bubble chamber, or a click localised in a detector. Measurement of position will be treated as fundamental in this paper, though it should be clear that the results will hold for statements about any family of measurements of uniform resolution based on a single property.

## 2   Many Valued Logic

We regard logic as the branch of mathematics in which we study relationships between specific types of statement. For example Aristotelian logic discusses basic types of logical statement such as the syllogism. At the beginning of this century philosophers and logicians became interested in the notion of 'ideal language' and propositional logic has been studied as a prototype. In classical, or crisp, logic, the truth of a proposition is given by the values 0 (definitely false) or 1 (definitely true). Often (but not always) the logical structure is set up in such a way that the truth value of a proposition corresponds to the actual truth of the proposition. A limitation of classical logic is that it uses only two truth values, whereas ordinary language is also capable of describing levels of certainty. Many valued logics were introduced in the 1920s for dealing with the intuitive idea of degrees of certainty, and key ideas in the use and interpretation of many valued logic were described by Max Black [7]. A survey of many valued logics has been prepared by Rescher [8].

Many valued logics actually predate the study of many valued logic by some margin. The most familiar is probability theory [9]. Probability theory was used (some have said abused) by the pioneering statistician Thomas Bayes and is the foundation for Bayesian estimation, or Bayesian reasoning. Bayesian reasoning may be criticised for using probabilistic law in situations where it does not strictly apply, but it is useful for approximation. A more modern and more general many valued logic is fuzzy logic, created by Professor Lofti Zadeh, [10][11] which has been used with considerable success in systems science for approximate reasoning based on imprecise information as is typically supplied by natural language, and has developed into a major subject area in its own right [12][13].

Another familiar many valued logic is found in bookmaking. The odds are often considered to reflect a probability that a given horse will come in first, though there is clearly no sample space on which probability can be defined. The odds are actually a product of the subjective valuations of the punters and the bookie. Nonetheless they are determined by bookmakers according to definite rules, and generate predictable results in that they ensure the bookie's long term income, so cannot be held to be purely subjective or arbitrary (the inherent bias reflects ordinary thought, and illustrates the distinction between a truth value and actual truth).

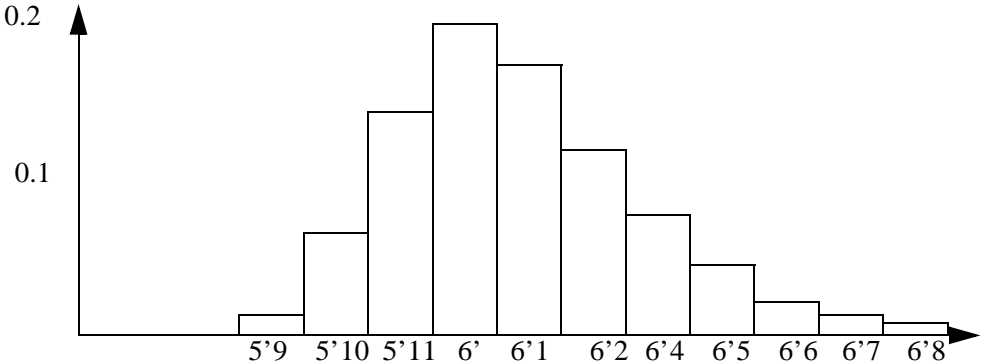

**Figure 1:** Possible approximate truth values for the statement "Joe Bloggs is a tall man", as might be used in fuzzy logic

In a typical many valued logic a real valued function, $f_P(x)$, is used as a measure of the certainty of the truth of a proposition $P[x]$. Propositions are given real truth values between 0 and 1, representing a subjective measure of certainty of the truth of the proposition. Thus in many valued logic, truth is typically a real function taking values between 0 and 1. Figure 1 gives a graph which could be considered to approximate the statement "Joe Bloggs is a tall man". By this graph, the truth value of the proposition "Joe Bloggs is 6'0" is about 0.2. This does not correspond the (unknown) objective truth of the same proposition, which is either true or false.

In crisp (two valued) logic the position of a point is a mapping from $\mathbb{N}^3$ to {0,1}. This mapping confers the truth value 0 or 1 on any statement of the position of the point, so each statement of position is certainly true or certainly false. In many valued logic, other truth values are possible. For example in probability theory and fuzzy logic, the position of a point is a mapping from a subset of $\mathbb{N}^3$ to the real interval, [0,1], and expresses the probability or the level of certainty of each proposition

2.1         $P[x]$ = "The particle would be found at $x$ if a measurement of position were done".

We aim to show that quantum logic [14] is simply the many valued logic appropriate to a general theory of the measurement of physical quantities. In quantum logic we use complex truth values. The justification for using a complex truth value is that mathematical logic does not describe objective reality, but rather describes the language with which we talk about it. Truth is here a mathematical value applied to a concept, and like $\sqrt{-1}$, has no direct physical meaning. Thus the position function of a particle has a truth value at each co-ordinate, but uncertainty in position does not imply that the particle is physically spread across co-ordinate space. Quantum logic takes into consideration the empirical principle that measurement gives imprecise information about objective reality. The position function is simply the set of truth values describing the potential for finding the particle at each position. Thus, the position function of a particle is defined as a mapping from $\mathbb{N}^3$ to $\mathbb{C}$ expressing the level of certainty that a measurement of position will produce a given result.

## 3    Measurement

When we carry out measurement we set up many repetitions of a system, and record the frequency of each result. Probability is simply a prediction of frequency, so a mathematical model of physics must generate a probability for each possible result. Experiments to determine the behaviour of matter are based on knowledge of the initial state and measurement of the final state. We require laws of physics to predict the change taking place between the first measurement and the second. There may be a practical difference between an initial measurement and a final one, but both are treated as simply measurements and described formally in the same way.

We have discarded the notion of a measurement *of* a property of a particle *by* an apparatus. Measurement is the generation of a value out of the combined interactions of apparatus, and does not necessarily imply that that value exists prior to measurement. Clearly we cannot carry out measurement of a particle in isolation, or measurement with an apparatus and no particle, so when we speak of performing a measurement by an apparatus on a particle we artificially separate the two parts of a physical process. Strictly a particle (or subsystem) under study and the apparatus used to study it are a single system consisting of many interacting particles. It is found that measurement of a property results in a definite value of that property, and this value is used to name the state of matter, particle and apparatus, which generated it.

**Definition:** The ket $|f\rangle$ is a label for a state of particle and apparatus, as categorised by the result, $f$, of measurement. A bra is an alternative representation of a ket.

Kets are labels, or names, associated with physical states. This is significant because when we introduce the laws of vector space (i.e. the principle of superposition), we will be speaking of the properties of a labelling system or naming convention, not of objective properties of matter. But, in keeping with common practice, we loosely refer to kets as states. Physically the state of a particle consists of its relationships to other particles, but we define a mathematical structure in which kets correspond to normal, but essentially inaccurate, use of language in which we talk of "the state of the particle". Correspondence with reality will be restored in the observable predictions of the theory.

**Definition:** The braket is the quantum logical truth value describing the degree of certainty that, if a suitable measurement were performed, the state labelled $|g\rangle$ would follow from the initial state $|f\rangle$

In particular the braket $\langle x|f\rangle$ is a measure of the level of certainty that, if it were performed, an experiment to measure position in state $|f\rangle$ would result in the value $x$. It is clear in this definition that the braket is not a quantity describing a real thing, but an invention based on supposition. According to the rules of many valued logic we have

3.1     $|\langle f|g\rangle| = 1$         if $f$ is certain to follow $g$

3.2     $|\langle f|g\rangle| = 0$         if $f$ cannot result from $g$

A central issue in the application of many valued logic is the determination of a truth function suited to the situation under consideration. We now seek further constraints on the braket.

**Figure 2**: Traditionally in quantum mechanics, kets have been thought of as describing the state of the particle, but what we actually measure is the state of the apparatus. That is to say we read the value of the state from the apparatus, and apply that value to the state of the particle. There is no fundamental difference between the matter in the apparatus and the matter being measured. In spite of the difference in the arrangement of the particles of matter constituting each, they are both labelled by bras and kets. By definition, if the state of the apparatus is named by a particular ket, the state of the particle is named by the same ket.

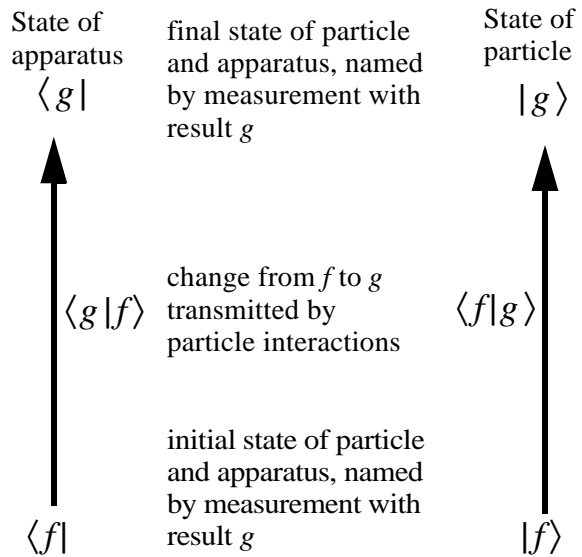

The particle alters the state of the apparatus, since the apparatus is designed to give a reading of the state of the particle. The apparatus also alters the state of the particle, since it is impossible to measure the particle without interacting with it. Whatever the actual configuration of matter, both state of particle and state of apparatus are categorised, or named, according to the same value derived from measurement, ensuring correspondence between the two. Because the labelling of the particle and the labelling of the apparatus are identical, we adopt a definition of a truth value such that the uncertainty in the apparatus is equal to the uncertainty in the position of the particle. We can regard the ket as labelling the state of the measuring apparatus and the bra as labelling the state of the particle, so that, if the apparatus is in the (known) state $|f\rangle$, then the particle is labelled by the bra $\langle f|$. Then, the degree of certainty for a transition of the apparatus to the state $\langle g|$ is $\langle g|f\rangle$, and the corresponding degree of certainty for the particle is $\langle f|g\rangle$. Uncertainty is divided equally between particle and apparatus, so a consistent definition of the braket is constrained to factorise probability

3.3         *Probability*($g$ leads to $f$) = $\langle f|g\rangle\langle g|f\rangle$

Probability is a real valued function so $\langle f|g\rangle = \overline{\langle g|f\rangle}$. 3.3 is a defining relationship for an artificial mathematical structure, not part of a description of physical reality.

It is worth remarking that the Copenhagen interpretation describes the particle with an uncertainty relation and the apparatus as certain. Here uncertainty is divided equally between particle and apparatus, but uncertainty in a macroscopic apparatus is governed by statistical law. Macroscopic phenomena are configurations of very large numbers of particles, so classical law consists of the average behaviour of large numbers of particles, each individually obeying quantum law.

## 4  Ket Space

In all practical measurements the apparatus has a finite resolution. Clearly the expression of the result of measurement in a real interval is equivalent to its expression as the centre point of that interval, together with knowledge of the resolution of the apparatus. In units of the resolution of the apparatus, the result of a measurement of position at a particular time is a point in $\mathbb{N}^3$. There is a practical bound on the magnitude of the result, so the result of measurement of position is in a finite region $N \subset \mathbb{N}^3$. N is not a bound on the universe and merely has to be large enough to be able to say with certainty that N contains any particle under study, i.e. the position function of the particle vanishes outside of N.

**Definition:** The coordinate system is $N = (-\nu, \nu] \otimes (-\nu, \nu] \otimes (-\nu, \nu] \subset \mathbb{N}^3$ for some $\nu \in \mathbb{N}$, where $(-\nu, \nu] = \{x \in \mathbb{N} | -\nu < x \leq \nu\}$.

**Definition:** Let $\chi$ be the magnitude of conventional units in terms of the resolution of the apparatus.

**Definition:** $\forall x \in N$, $|x\rangle$ is the ket denoting a measurement of position $x$. $|x\rangle$ is called a position ket.

**Definition:** Let $\mathbb{H}_0$ be the set of kets resulting from a measurement of position of an elementary particle in N. $\mathbb{H}_0$ contains kets for all physically realised measurements of position, but also kets for measurements of position which may be made in principle, and it also contains kets which may not be realised either in principle or in practice.

**Definition:** Construct a vector space, $\mathbb{H}$, over $\mathbb{C}$, with basis $\mathbb{H}_0$. $\mathbb{H}$ can be represented as the set of $2\nu \times 2\nu \times 2\nu \times 1$ matrices generated by the operations of addition and multiplication by a scalar from basis kets represented by a matrix containing one 1 and all other entries equal to 0.

To make this concrete, first lets consider a simplified example of a measurement which has only four possible outcomes, 1, 2, 3 and 4. Then the "co-ordinate system", N, is one dimensional and $N = \{1, 2, 3, 4\}$. It immediate from the definition that $\mathbb{H}_0 = \{|1\rangle, |2\rangle, |3\rangle, |4\rangle\}$ and in the matrix representation

$$4.1 \qquad |1\rangle = \begin{bmatrix} 1 \\ 0 \\ 0 \\ 0 \end{bmatrix}, \qquad |2\rangle = \begin{bmatrix} 0 \\ 1 \\ 0 \\ 0 \end{bmatrix}, \qquad |3\rangle = \begin{bmatrix} 0 \\ 0 \\ 1 \\ 0 \end{bmatrix}, \qquad |4\rangle = \begin{bmatrix} 0 \\ 0 \\ 0 \\ 1 \end{bmatrix}$$

and $\mathbb{H}$ is the space of $4 \times 1$ matrices which can be composed from the elements of $\mathbb{H}_0$ by the operations of addition and scalar multiplication. In general a $4 \times 1$ matrix does not describe a particular state of the system, but it is a property of a state in which a measurement may have any of the four possible outcomes, and we use this property to name the state and to calculate the probability of each of those outcomes. The naming convention is not assumed to be unique. The same $4 \times 1$ matrix may apply to different configurations of matter, and different $4 \times 1$ matrices may convey the same information about the state.

Vector space introduces intuitive logical operations between uncertain propositions. Addition corresponds to logical OR, and multiplication by a scalar gives an intuitive idea of weighting due to the level of certainty in each option given to logical OR. This is justified because kets are simply labels for possible states of matter, not descriptions of reality. Thus the principle of superposition is a definitional truism, not a physical assumption. Vector space extends the labelling system from $\mathbb{H}_0$ to $\mathbb{H}$. Multiplication by a scalar only has logical meaning as a weighting between alternatives, $\forall |f\rangle \in \mathbb{H}$, $\forall \lambda \in \mathbb{C}$ such that $\lambda \neq 0$, so $\lambda |f\rangle$ is a label for the same physical state as $|f\rangle$. We can therefore renormalise kets as we choose, without affecting their use as labels for states.

In a measurement of position a particle can be found anywhere, but it is only found in one place at a time. The braket, or position function, which describes this is a Kronecker delta, renormalised to

4.2 $\qquad \forall \boldsymbol{x}, \boldsymbol{y} \in \mathrm{N}, \langle \boldsymbol{x}|\boldsymbol{y}\rangle = \chi^3 \delta_{xy}$

**Definition:** With this normalisation, the position function of a particle in the state $|f\rangle \in \mathbb{H}$ is the function $\mathrm{N} \to \mathbb{C}$ defined by

4.3 $\qquad \forall \boldsymbol{x} \in \mathrm{N}, \boldsymbol{x} \to \langle \boldsymbol{x}|f\rangle$

From the property that any vector can be expanded in terms of a basis we have

$$\forall |f\rangle \in \mathbb{H}, \exists f: \mathrm{N} \to \mathbb{C} \text{ such that } |f\rangle = \sum_{\boldsymbol{x} \in \mathrm{N}} \frac{1}{\chi^3} f(\boldsymbol{x})|\boldsymbol{x}\rangle$$

By applying $\langle \boldsymbol{x}|$ to both sides and using 4.2 we have $f(\boldsymbol{x}) = \langle \boldsymbol{x}|f\rangle$, so

4.4 $\qquad \forall |f\rangle \in \mathbb{H}, |f\rangle = \sum_{\boldsymbol{x} \in \mathrm{N}} \frac{1}{\chi^3}|\boldsymbol{x}\rangle\langle \boldsymbol{x}|f\rangle$

So the braket is given by the hermitian form known as the scalar product, i.e.

4.5 $\qquad \langle g|f\rangle = \sum_{\boldsymbol{x} \in \mathrm{N}} \frac{1}{\chi^3}\langle g|\boldsymbol{x}\rangle\langle \boldsymbol{x}|f\rangle$

There is a homomorphic correspondence between $\mathbb{H}$ and the space of complex functions on N given by the correspondence between a ket and its position function. Position function can also be regarded as the set of components of a vector in a particular basis.

## 5   Momentum Space

**Definition:** Momentum space is $\mathrm{M} = (-\pi, \pi] \otimes (-\pi, \pi] \otimes (-\pi, \pi]$ where $(-\pi, \pi]$ is the real interval $\{x \in \mathbb{R} | -\pi < x \leq \pi\}$. Elements of momentum space are called momenta.

**Definition:** For each value of momentum $\boldsymbol{p} \in \mathrm{M}$, define a ket $|\boldsymbol{p}\rangle$, known as a plane wave state, by the position function

5.1 $\qquad \langle \boldsymbol{x}|\boldsymbol{p}\rangle = \left(\frac{\chi}{2\pi}\right)^{\frac{3}{2}} e^{-i\boldsymbol{x}\cdot\boldsymbol{p}}$

With this definition of momentum the origin of the uncertainty principle is immediate, since by definition of a state of definite momentum is a composition of states of different positions. Its relationship to classical momentum is found from the statistical analysis of the behaviour of many particles [16].

**Definition:** For each ket $|f\rangle$ define the momentum space function $F(\boldsymbol{p}) = \langle \boldsymbol{p}|f\rangle$

Then, by 4.5, $F$ can be expanded as a trigonometric polynomial

5.2 $\qquad \langle \boldsymbol{p}|f\rangle = \left(\frac{\chi}{2\pi}\right)^{\frac{3}{2}} \sum_{\boldsymbol{x} \in \mathrm{N}} \frac{1}{\chi^3}\langle \boldsymbol{x}|f\rangle e^{i\boldsymbol{x}\cdot\boldsymbol{p}}$

Clearly the cardinality of the plane wave states is greater than the cardinality of $\mathbb{H}_0$, so plane waves are not a basis. But the position function can be found in terms of plane waves by Fourier analysis

5.3 $\qquad \langle \boldsymbol{x}|f\rangle = \left(\frac{\chi}{2\pi}\right)^{\frac{3}{2}} \int_\mathrm{M} d^3\boldsymbol{p} \langle \boldsymbol{p}|f\rangle e^{-i\boldsymbol{x}\cdot\boldsymbol{p}}$

The lack of symmetry between momentum space and co-ordinate space reflects the idea that position is closely associated with the fundamental nature of matter, whereas momentum is a human construction.

The dependency of momentum space functions on N expresses a bound on the magnitude of momentum which could be measured by a given apparatus. In a general theory of measurement, the value χ depends on the resolution of the apparatus (taken to be a constant), not on the fundamental structure of matter, showing that vector space is appropriate to a general theory of measurement, not just quantum mechanics. Then using the definition of momentum by Fourier series, the uncertainty principle follows directly, as was first proved formally by Robertson [17]. But instead of a dependency on Planck's constant, uncertainty depends on the finite resolution of the apparatus. In an optimally accurate measurement, the resolution of the apparatus depends on the fundamental structure of matter. The appearance of vector space and Planck's constant indicates the ultimate resolution of any physical measurement, and hence that Planck's constant is a measure of the scale of the fundamental structure of matter.

## 6  Classical Correspondence

In a measurement of position, the ket, $|f\rangle$, naming the initial state of the apparatus is changed into a ket named by a position in $X$, where $X$ is a region of space (i.e. a set of possible results of measurement) determined by the measuring apparatus. The operator effecting the change is

6.1 $$Z(X) = \sum_{x \in X} \frac{1}{\chi} |x\rangle\langle x|$$

as is shown by direct application

6.2 $$Z(X)|f\rangle = \sum_{x \in X} \frac{1}{\chi} |x\rangle\langle x|f\rangle$$

since the resulting state is a weighted logical or between positions in $X$. Applying $Z$ a second time to 6.2

$$Z(X)Z(X)|f\rangle = \sum_{y \in X} \frac{1}{\chi}|y\rangle\langle y| \sum_{x \in X} \frac{1}{\chi}|x\rangle\langle x|f\rangle$$

$$= \sum_{y \in X} \frac{1}{\chi}|y\rangle\langle y|f\rangle \qquad \text{by 4.2}$$

So $Z(X)$, is a projection operator

6.3 $$Z(X)Z(X) = Z(X) \qquad \text{by 6.1}$$

reflecting the observation that a second measurement of a quantity gives the same result as the first (e.g. [18]). By applying 6.3 to the state $|f\rangle$ normalised so that $\langle f|f\rangle = 1$ obtain

6.4 $$\langle f|Z(X)Z(X)|f\rangle = \sum_{x \in X} \frac{1}{\chi}\langle f|x\rangle\langle x|f\rangle$$

But 6.4 is the sum of the probabilities that the particle is found at each individual position, $x \in X$. In other words it is the probability that a measurement of position finds the particle in the region $X$. In the case that $X$ contains only the point $x$, $X = \{x\}$, 6.2 becomes

6.5 $$Z(x)|f\rangle = \frac{1}{\chi}|x\rangle\langle x|f\rangle$$

Thus, the position function, $\langle x|f\rangle$, can be reinterpreted as the magnitude of the projection, $Z(x)$, from the state $|f\rangle$ of the apparatus into the state $|x\rangle$.

$Z(X)$, is not simply a mathematical device to produce a result; it actually summarises the physical processes taking place in the interactions involved in a measurement of position. If a measurement of position performed on the state $|f\rangle$ has resulted in a position in $X$, $Z(X)$ has, in effect, been applied to $|f\rangle$ and it is assumed that $Z(X)$ is generated by some combination of particle interactions. Classical probability theory describes situations in which every parameter exists, but some are not known. Probabilistic results come from different values taken by unknown parameters. We have a similar situation here. There are no relationships between particles apart from those generated by interactions. An experiment can be described by a large configuration of particles incorporating the measuring apparatus as well as the process being measured. The configuration of particles has been largely determined by setting up the experimental apparatus, but the precise pattern of interactions is unknown. Thus the current model describes a classical probability in which the unknowns lie in the configuration of interacting particles.

The probability that the interactions combine to $Z(x)$ is given by

6.6 $$\langle f|Z(x)Z(x)|f\rangle = \langle f|Z(x)|f\rangle = \frac{1}{\chi}\langle f|x\rangle\langle x|f\rangle = \frac{1}{\chi}|\langle x|f\rangle|^2$$

Thus, 3.3 can be understood as a classical probability function, where the variable, $x$, runs over the set of projection operators,

6.7 $$Z(x) = \frac{1}{\chi}|x\rangle\langle x|$$

such that each $Z(x)$ is generated by an unknown configuration of particle interactions in measurement.

In general, measurements generate numerical values and are repeated many times over from the same starting state. Then the average value of the result is taken. Expectation is the term used in statistics for the prediction of an average value. Under the laws of statistics, the more repetitions, the closer the average value will be to the expectation of the measurement. By the definition of expectation in statistics, if $O(x)$ is a real function of a position, $x \in \mathbb{N}$, then, given the state $|f\rangle \in \mathbb{H}$ $x$ is a random variable with probability function $|\langle f|x\rangle|^2$. and the expectation of $O(x)$ is

6.8 $$\langle O\rangle = \sum_{x \in X}\frac{1}{\chi}\langle f|x\rangle O(x)\langle x|f\rangle$$

It is straightforward to generalise the analysis to let $O(x)$ be a real valued functional. If we define an operator on $\mathscr{F}$ by the formula

6.9 $$O = \sum_{x \in X}\frac{1}{\chi}|x\rangle O(x)\langle x|$$

then the expectation of $O$ given the initial state $|f\rangle \in \mathbb{H}$ is

6.10 $$\langle O\rangle = \langle f|O|f\rangle \qquad\qquad\qquad\qquad\qquad\qquad\qquad\text{by 6.8 and 6.9}$$

$O$ is linear over independent multiparticle states, so 6.8 applies also to the expectation for all $|f\rangle \in \mathscr{F}$. $O$ is hermitian, so there is a particular class of kets, called eigenkets, such that if $|f\rangle$ is an eigenket, then $\exists r \in \mathbb{R}$ known as the eigenvalue associated with $|f\rangle$ such that

6.11 $$O|f\rangle = r|f\rangle.$$

If all physical processes are described by a composition of interaction operators then the existence of an observable quantity depends not on whether an observation takes place, but on the configuration of matter. If the interaction operators describing a physical process combine to generate a hermitian operator, then the corresponding observable quantity exists, independent of observation or measurement. The state is said to be an eigenstate of the observable, and is labelled by an eigenket of the operator. The value of the observable quantity is given by the corresponding eigenvalue

6.12 $$\langle O \rangle = \langle f|O|f \rangle = \langle f|r|f \rangle = r\langle f|f \rangle = r$$

We know from experiment that measurements generate definite results, and thereby provide definite categorisations of states by means of a kets. This is equivalent to the application of a projection operator. In a statistical analysis of a large number of particles, each result names a physical process described by a combination of interaction operators equivalent to a projection operator. Under the identification of addition with logical OR the expectation of all the results is a hermitian operator equal to a weighted sum over a family of projection operators. Classical laws are derived from the expectation, 6.12, of the interactions of large numbers of particles.